\documentclass[%
 reprint,
amsmath,amssymb,
aps,
prl,
]{revtex4-2}

\usepackage{graphicx}
\usepackage{dcolumn}
\usepackage{bm}
\usepackage{hyperref}
\hypersetup{%
 setpagesize=false,%
 bookmarks=false,%
 hypertexnames=false,
 colorlinks=true,%
 allcolors={blue},
 pdftitle={},%
 pdfsubject={},%
 pdfauthor={},%
 pdfkeywords={}}

\usepackage{physics}

\usepackage[normalem]{ulem}
\usepackage{xcolor}

\newcommand{\nn}{\nonumber \\}
\newcommand{\vk}{\bm{k}}
\newcommand{\vq}{\bm{q}}
\newcommand{\vA}{\bm{A}}

\DeclareRobustCommand{\MCC}{\bgroup\markoverwith{\textcolor{magenta}{\rule[.5ex]{2pt}{0.6pt}}}\ULon}

\begin{document}

\preprint{APS/123-QED}

\title{Orbital FFLO and layer-selective FFLO phases in trilayer NbSe$_2$}%

\author{Michiya Chazono}
\email[]{mcphys77@gmail.com}
\affiliation{Department of Physics, Graduate School of Science, Kyoto University, Kyoto 606-8502, Japan}

\author{Youichi Yanase}%
\email[]{yanase@scphys.kyoto-u.ac.jp}
\affiliation{Department of Physics, Graduate School of Science, Kyoto University, Kyoto 606-8502, Japan}

\date{\today}

\begin{abstract}
Finite-momentum superconductivity has become an important research topic in condensed matter physics. In particular, the orbital Fulde-Ferrell-Larkin-Ovchinnikov (FFLO) state, which is stabilized in atomically thin films by the orbital effect of an external magnetic field, has been getting attention as a fascinating finite-momentum superconducting state recently. We study the phase diagram of the trilayer Ising superconductor NbSe$_2$ in the in-plane magnetic field, taking into account the orbital effect, the paramagnetic effect, and the spin-orbit coupling. The finite-momentum gap structure in the high-field region is shown by a large-scale numerical calculation based on the Bogoliubov-de Gennes equation. We find an exotic superconducting phase, 
a layer-selective FFLO phase, in which finite-momentum Cooper pairs coexist with zero-momentum Cooper pairs, separated from the orbital FFLO phase.    

\end{abstract}

\maketitle


\emph{Introduction.}--- While the center-of-mass momentum of Cooper pairs vanishes in conventional superconducting states, Cooper pairs with finite total momentum can be stabilized under various conditions~\cite{FF1964, LO1965_Eng, Bauer2012, Smidman2017, Kittaka2023, Asaba2024, Shimahara1996, Kanasugi2022, Chazono2023, Kitamura2023, Kitamura2022, Shaffer2023, Ticea2024, Velasco2023, Setty2023, Sumita2016,Sumita2017,Sumita2023, Zhang2024AM, Mootz2024, Clepkens2024, Sboychakov2025}. Such finite-momentum superconductivity has recently received much attention, for instance as platforms of the superconducting diode effect~\cite{Ando2020, Narita2022, Lin2022, Pal2022, Levichev2023, Yuan2022, Daido2022, He2022, Scammell2022, Ilic2022, Legg2022, Daido2022-2, Ikeda2022, He2023, Nunchot2024, Nakamura2024, Zhang2024FP, Daido2024, Matsumoto2024,Nagaosa-Yanase,Nadeem2023}. The Pauli paramagnetic effect plays a central role in typical finite-momentum superconducting states in an external magnetic field, such as the Fulde-Ferrell-Larkin-Ovchinnikov (FFLO) state~\cite{FF1964, LO1965_Eng} and the helical superconducting state in Rashba superconductors~\cite{Bauer2012, Smidman2017}. On the other hand, the orbital effect is now attracting much attention as the origin of the so-called \textit{orbital} FFLO state, in which Cooper pairs acquire finite momentum due to the orbital-effect-induced Fermi surface shift~\cite{Wan2023, Zhao2023, Cho2023, Yang2024, Cao2024, Ji2024, 
Zhao2024, Watanabe2015, Liu2017, Nakamura2017, Masutomi2020, Qiu2022, Xie2023, Yuan2023, Nag2024, Yan2024, Yuan2025, Zhu2025}.

Transition metal dichalcogenides (TMDs) are promising candidates for the orbital FFLO superconductors. The noncentrosymmetric crystal structure of the monolayer TMDs induces the Ising-type spin-orbit coupling (SOC) in each layer, which locks electron spins in an out-of-plane direction. The SOC suppresses the Pauli depairing effect of the in-plane magnetic field and leads to the Ising superconductivity~\cite{Xi2016, Saito2016} with extraordinary high critical fields. Therefore, the orbital effect can be dominant in superconducting multilayer TMDs, and the orbital FFLO states can be realized~\cite{Liu2017, Nakamura2017, Xie2023, Yuan2023, Nag2024, Yan2024, Yuan2025, Zhu2025}. Several experiments have shown the orbital FFLO states in various TMDs from bilayer to bulk systems~\cite{Wan2023, Zhao2023, Cho2023, Yang2024, Cao2024, Ji2024, 
Zhao2024}.

Although the orbital FFLO states in bilayer and other even-layer TMD systems have been intensively investigated theoretically~\cite{Liu2017, Nakamura2017, Xie2023, Yuan2023, Nag2024, Yan2024, Yuan2025, Zhu2025},
odd-layer systems have received less attention in this context. Meanwhile, a recent experiment reported multiple superconducting phases and suggested a finite-momentum superconducting state in the trilayer NbSe$_2$ segments in the misfit compound (PbSe)$_{1.14}$(NbSe$_2$)$_3$~\cite{Itahashi2024}. Therefore, a theoretical analysis of superconductivity in odd-layer TMDs is urgently required.


In this paper, 
we theoretically investigate the superconducting phase diagram of the trilayer NbSe$_2$ in the in-plane magnetic field, as a fundamental example of odd-layer TMDs. The superconducting gap structure is calculated by solving the Bogoliubov-de Gennes (BdG) equation based on a minimal model of the trilayer NbSe$_2$. 
Clarifying the interplay of the orbital effect, the Pauli paramagnetic effect, and the Ising spin-orbit coupling, 
we show multiple finite-momentum superconducting states in the high-magnetic-field region. In particular, an exotic superconducting phase, called the layer-selective FFLO phase, where finite-momentum and zero-momentum Cooper pairs coexist, is predicted. We clarify the mechanism of this state and highlight the properties qualitatively different from the orbital FFLO state.

\begin{figure}[t]
  \centering
  \includegraphics[width=60mm]{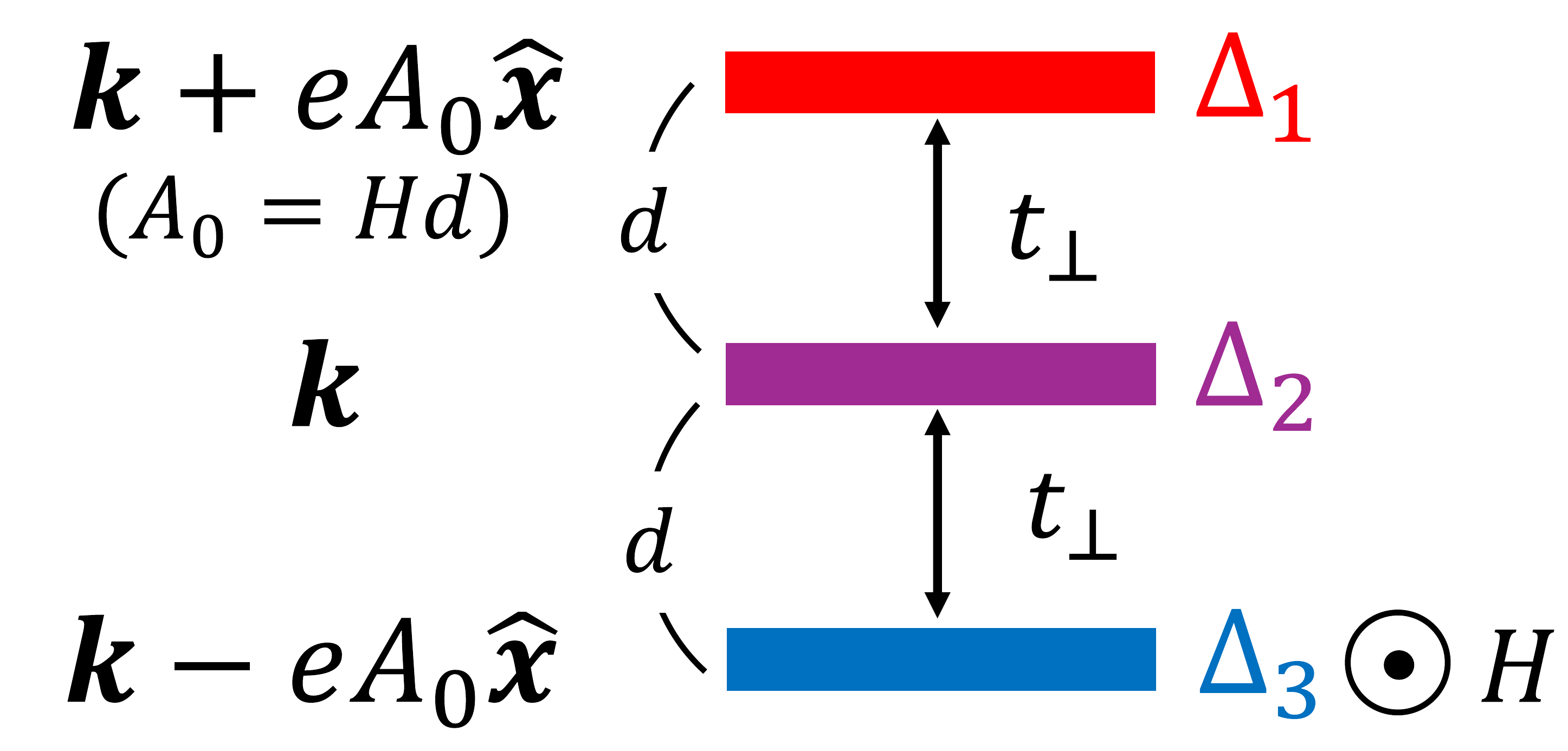}
  \caption{ 
  Schematic image of the model Eq.~\eqref{Hamiltonian}. Each layer represents a monolayer NbSe$_2$ with the Ising SOC 
  and the three layers are coupled by the interlayer hopping $t_\perp$.
 }
\label{fig:Model_schematic}
\end{figure}

\emph{Model.}--- We consider the in-plane magnetic field 
in the $y$-direction: $\bm{H} = H\hat{y}$, and the corresponding vector potential is chosen as $\vA = Hz\hat{x}$. To study the superconducting phase diagram we adopt the following minimal model of the trilayer 2H-type NbSe$_2$
(its minimal schematic is shown in Fig.~\ref{fig:Model_schematic})
:
\begin{align}
\mathcal{H}
= \mathcal{H}_\textrm{kin}
+ \mathcal{H}_\textrm{Ising} 
+ \mathcal{H}_\perp
+ \mathcal{H}_\textrm{Zeeman}
+ \mathcal{H}_\textrm{int},
\label{Hamiltonian}
\end{align}
where
$\mathcal{H}_\textrm{kin} =\sum_{m=1}^3 \sum_{\vk s }\xi(\vk) c^\dag_{\vk s m} c_{\vk s m}$ is the intra-layer hopping term, and
$\mathcal{H}_\textrm{Ising} = \sum_{m=1}^3 \sum_{\vk s s^\prime} (-1)^{m-1} \bm{g}(\vk) \cdot \bm{\sigma}_{s s^\prime} c^\dag_{\vk s m} c_{\vk s^\prime m}$ is the layer-dependent Ising-type SOC term arising from the crystal structure of 2H-type NbSe$_2$. The parameters of these terms are chosen to reproduce the low-energy band dispersion of the monolayer NbSe$_2$ obtained by \textit{ab initio} calculations~\cite{Sticlet2019,Supplemental}. 
We introduce the inter-layer coupling between the adjacent layers:
$\mathcal{H}_\perp = t_\perp \sum_{\vk s \langle m m^\prime \rangle} c^\dag_{\vk s m} c_{\vk s m^\prime}.$ For the influence of the magnetic field, we consider both the Pauli paramagnetic effect and the orbital effect. The former gives the Zeeman field:
$\mathcal{H}_\textrm{Zeeman} = - \mu_{\rm B} \sum_{m=1}^3 \sum_{\vk s s^\prime} \bm{H} \cdot \bm{\sigma}_{s s^\prime} c^\dag_{\vk s m} c_{\vk s^\prime m}$, and the latter is taken into account by the minimal coupling with replacement $\vk \rightarrow \vk + e\vA^{(m)}$ in $\mathcal{H}_\textrm{kin}$ and $\mathcal{H}_\textrm{Ising}$,
where $\vA^{(m)}$ is the layer-dependent vector potential. Regarding the middle layer as the origin ($z=0$), the vector potential is given as $\vA^{(1)} = A_0\hat{x},~ \vA^{(2)} = \bm{0},$ and $\vA^{(3)} = -A_0\hat{x}$ on the top, middle, and bottom layers, respectively. Here, $A_0 = H d$ with the distance between adjacent layers $d$. 
We assume the onsite attractive interaction and adopt the mean-field approximation for simplicity:
\begin{align}
\mathcal{H}_\textrm{int} 
&= - \frac{U}{N} \sum_{m=1}^3 \sum_{i} c^\dag_{i \uparrow m} c^\dag_{i \downarrow m} c_{i \downarrow m} c_{i \uparrow m} \nn
& \approx \sum_{m=1}^3 \sum_{\vk, \vq} \left(\Delta_m (\vq) c^\dag_{\vk + \vq/2 \uparrow m} c^\dag_{-\vk + \vq/2 \downarrow m} + h.c. \right) \nn
& \hspace{16pt} + \frac{N}{U} \sum_{m=1}^3 \sum_{\vq} \abs{\Delta_m (\vq)}^2.
\label{Hint}
\end{align}
Since the orbital effect induces Fermi surface shift along the $x$-direction and stabilizes the finite-momentum pairing, we assume that Cooper pairs' total momentum $\vq$ is along the $x$-direction: $\vq = q\hat{x}$ and the superconducting state is uniform in the $y$-direction. 

\emph{Method.}---
We 
obtain the phase diagram in two steps. First, to roughly estimate the superconducting phase diagram taking into account the finite-momentum pairing, we solve the linearized gap equation assuming the Fulde-Ferrell-type single-$\vq$ gap functions $\Delta_m (\vq) = \Delta_m e^{i \vq \cdot \bm{r}}$:
\begin{align}
1 - U \chi^0 (T,H,\vq) = 0,
\label{LGE}
\end{align}
where $\chi^0 (T,H,\vq)$ is the irreducible superconducting susceptibility matrix with total momentum $\vq$:
\begin{align}
& \left[ \chi^0 (T,H,\vq) \right]_{m m^\prime} \nn
&= - T \sum_{\vk l}  \left[ G^{\uparrow \uparrow}_{m  m^\prime} (\vk+\vq/2, i\omega_l) G^{\downarrow \downarrow}_{m m^\prime} (-\vk+\vq/2, -i\omega_l) \right. \nn
& \hspace{0.4cm} \left. - G^{\uparrow \downarrow}_{m  m^\prime} (\vk+\vq/2, i\omega_l) G^{\downarrow \uparrow}_{m m^\prime} (-\vk+\vq/2, -i\omega_l) \right].
\label{chi0}
\end{align}
For given $H$ and $\vq$, the superconducting transition temperature $T_{\rm c}(H,\vq)$ satisfies 
\begin{align}
\lambda_\textrm{max} \left[ \chi^0 (T_{\rm c}(H,\vq),H,\vq) \right] = 1/U,
\label{Tc_def}
\end{align}
where $\lambda_\textrm{max} [\chi^0 (T,H,\vq)]$ is the maximum eigenvalue of $\chi^0 (T,H,\vq)$. The most stable Cooper pairs' momentum $\vq_0$ is determined as
\begin{align}
T_{\rm c}(H,\vq_0) = \max_{\vq} \left[ T_{\rm c}(H,\vq) \right].
\label{q0_def}
\end{align}
We denote the transition temperature of the zero-momentum Ising superconducting state as $T_{\rm c}^{\rm Ising}(H)=T_{\rm c}(H,0)$ for later use. The real transition temperature can be higher than $T_{\rm c}^{\rm Ising}(H)$ when Cooper pairs have finite momentum.

Second, to determine the superconducting state, we consider multiple-$\bm q$ superconducting states, which are spatially inhomogeneous in general.
For this purpose, we solve the BdG equation in real space. Since the superconducting state is assumed to be uniform in the $y$-direction, the gap equation can be written in the following form~\cite{Supplemental}: 
\begin{align}
\Delta_{m}(i_x) = 
- \frac{U}{N_y} \sum_{k_y} \left\langle c_{i_x -k_y \downarrow m} c_{i_x k_y \uparrow m} \right\rangle.
\label{GE}
\end{align}
For 
large-scale numerical calculations, we use the Chebyshev polynomial method, which does not require diagonalization and is suitable for large system size calculations~\cite{Weisse2006, Covaci2010, Nagai2012, Nagai2020}:
\begin{align}
&\left\langle c_{i_x -k_y \downarrow m} c_{i_x k_y \uparrow m} \right\rangle 
\approx \sum_{n = 0}^{N_{\rm c}} g_n \mathcal{T}_n  \bm{e}(i_x, \uparrow, m)^T \bm{h}_n(i_x, \downarrow, m),
\label{SCMF_CPM}
\end{align}
where 
\begin{gather}
\bm{h}_0(i_x, s, m) = \bm{h}(i_x, s, m),~~
\bm{h}_1(i_x, s, m) = \tilde{\hat{H}} \bm{h}(i_x, s, m), \nn
\bm{h}_{n+1}(i_x, s, m) = 2 \tilde{\hat{H}} \bm{h}_n(i_x, s, m) - \bm{h}_{n-1}(i_x, s, m),
\label{hn_CPM} \\
\mathcal{T}_n = \frac{2}{1+\delta_{n,0}} \int^1_{-1} d\tilde{\omega} \frac{f(a \tilde{\omega} + b) T_n(\tilde{\omega})}{\pi \sqrt{1-\tilde{\omega}^2}},
\label{Tn_CPM}
\end{gather}
with the Chebyshev polynomial of the first kind $T_n (\omega)$, the Fermi distribution function $f(\omega)$ and the rescaled BdG Hamiltonian matrix $\tilde{\hat{H}} = (\hat{H} - b)/a$. The unit vector $\bm{e}(i_x, s, m)$ [$\bm{h}(i_x, s, m)$] is finite only at the element corresponding to the electron [hole] part characterized by the site $i_x$, spin $s$, and layer $m$. We adopt the Jackson kernel $g_n$ for better convergence and choose a sufficiently large cutoff $N_{\rm c} = 10000$. We impose a periodic boundary condition for the $x$-direction with 400 sites and for the $y$-direction with 100 sites. We confirmed the validity by reproducing the uniform gap function obtained by the momentum-space calculation~\cite{Supplemental}. 

We can also calculate the local density of states (LDOS) using the Chebyshev polynomial method:
\begin{align}
N_{i_x, s, m} (\omega)
\approx \sum_{n = 0}^{N_{\rm c}} g_n \mathcal{T}^\prime_n (\tilde{\omega}) \bm{e}(i_x, s, m)^T \bm{e}_n(i_x, s, m),
\label{LDOS_CPM}
\end{align}
where
\begin{gather}
\bm{e}_0(i_x, s, m) = \bm{e}(i_x, s, m),~~
\bm{e}_1(i_x, s, m) = \tilde{\hat{H}} \bm{e}(i_x, s, m), \nn
\bm{e}_{n+1}(i_x, s, m) = 2 \tilde{\hat{H}} \bm{e}_n(i_x, s, m) - \bm{e}_{n-1}(i_x, s, m),
\label{en_CPM} \\
\mathcal{T}^\prime_n (\tilde{\omega}) = \frac{2}{a (1+\delta_{n,0})} \frac{T_n(\tilde{\omega})}{\pi \sqrt{1-\tilde{\omega}^2}},
\label{Tnprime_CPM}
\end{gather}
with the rescaled energy $\tilde{\omega} = (\omega- b)/a$.

\begin{figure}[tb]
  \centering
  \includegraphics[width=85mm]{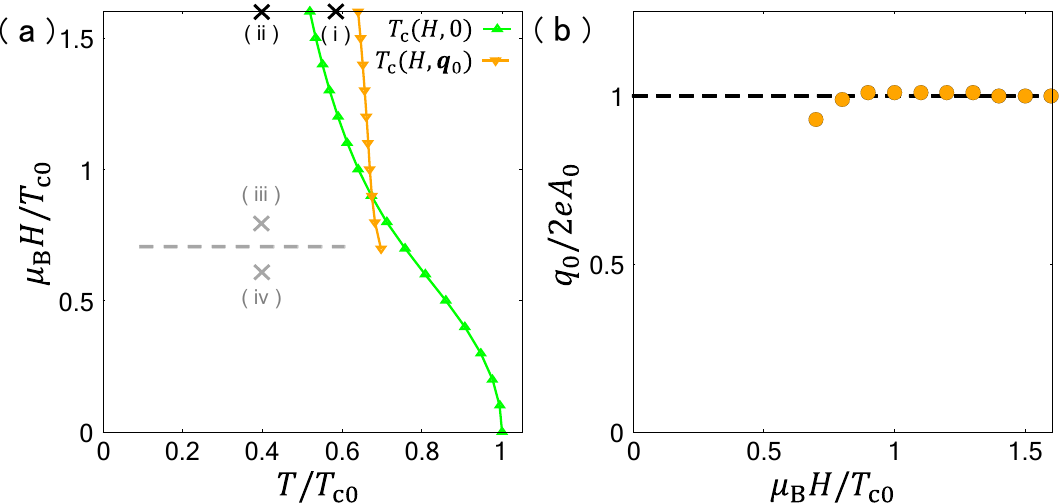}
  \caption{(a) Superconducting phase diagram obtained by the linearized gap equation. The green and yellow lines show the transition lines of the zero- and finite-momentum superconducting states, respectively. 
  The phase transition between the finite-momentum superconductivity and the uniform Ising superconductivity is expected to occur near the dashed line, and it is confirmed by solving the BdG equation. (b) Magnetic field dependence of the Cooper pairs' momentum $\abs{q_0}$ at the transition temperature. 
 }
\label{fig:PD_LGE}
\end{figure}

\emph{Results.}--- 
First, we show the superconducting phase diagram obtained by solving the linearized gap equation. Figure~\ref{fig:PD_LGE}(a) shows that a finite-momentum pairing state has the highest transition temperature in the high-field region. The superconducting states with Cooper pairs' momentum $+q_0 \hat{x}$ and $-q_0 \hat{x}$ have the same transition temperature due to the symmetry condition. The most stable momentum satisfies $\abs{q_0} \approx 2eA_0$ as shown in Fig.~\ref{fig:PD_LGE}(b), coinciding with the magnitude of the Fermi surface shift by the orbital effect on the outer layers. This indicates that the finite-momentum pairing is stabilized by the orbital effect rather than the paramagnetic effect~\cite{Liu2017, Nakamura2017, Xie2023}, supporting the orbital FFLO state. The $q_0$ shows a jump from $q_0=0$ to $\abs{q_0} \approx 2eA_0$ suggesting the first-order phase transition from the uniform Ising superconducting phase in the low-field region to the finite-momentum superconducting phase, which is confirmed by solving the BdG equation later. A similar first-order transition has been reported in NbSe$_2$ with intermediate thickness~\cite{Cao2024}.

The relation between the Cooper pairs' momentum and the layer-dependent gap function is summarized in Table~\ref{Table:LGE_q0_gap}.
In the solution for the orbital FFLO state with $q=\abs{q_0}$ ($q=-\abs{q_0}$), the gap function is largest on the top (bottom) layer. 
This is reasonable because the orbital effect shifts momentum on the top (bottom) layer in the positive (negative) direction and favors finite-momentum pairing in the same direction. 
Because the orbital effect does not shift momentum 
in the middle layer for our gauge choice, $\Delta_2$ is largest in the Ising superconducting state with $q = 0$. 
All of these features are consistent with the finite-momentum superconducting state induced by the orbital effect. 

\begin{table}[t]
\centering
\begin{tabular}{c|c|c} 
 $q > 0$ & $q = 0$ & $q < 0$ \\ \hline
 $\Delta_1 \gg \Delta_2 \gg \Delta_3$ & $\Delta_2 \gg \Delta_1 = \Delta_3$ & $\Delta_3 \gg \Delta_2 \gg \Delta_1$ \\ 
\end{tabular}
\caption{Relation between Cooper pairs' momentum $q$ and layer dependence of the gap function.} 
\label{Table:LGE_q0_gap}
\end{table}

\begin{figure}[tb]
  \centering
  \includegraphics[width=85mm]{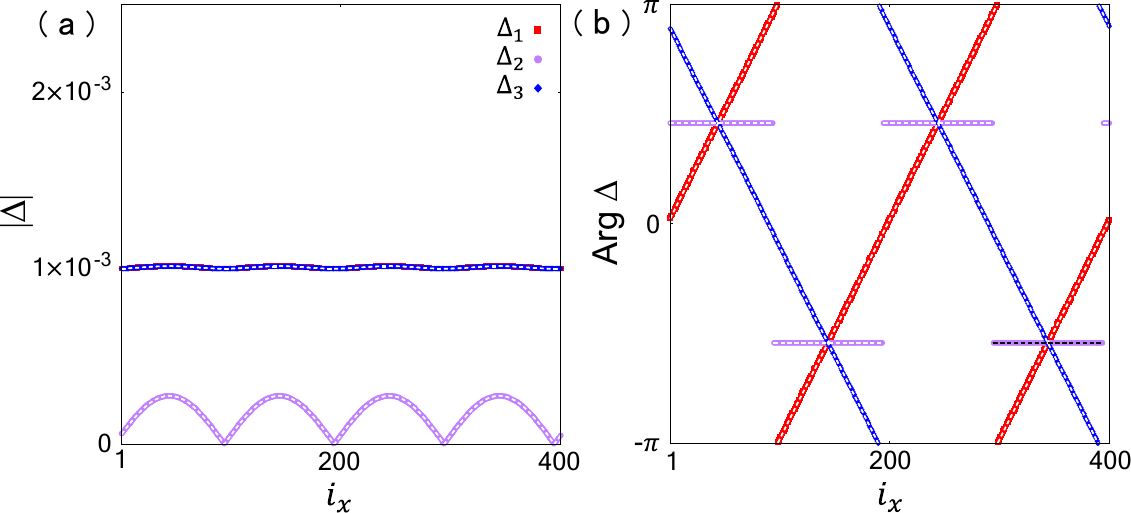}
  \caption{Spatial profile of (a) the amplitude and (b) the phase of the gap functions at $\mu_{\rm B} H_y/ T_{\rm c0} = 1.6$ and $T / T_{\rm c0} = 0.58$. The red, purple, and blue symbols show $\Delta_1(x)$, $\Delta_2(x)$, and $\Delta_3(x)$ on the top, middle, and bottom layers, respectively. The dashed lines are the fitting by Eq.~\eqref{Gap_form}. The amplitudes on the two outer layers are almost equivalent. 
 }
\label{fig:BdG_OFFLO}
\end{figure}

Next, we show the real-space gap structure obtained by solving the BdG equation with the Chebyshev polynomial method. Figure~\ref{fig:BdG_OFFLO} shows the amplitude and phase of the gap functions 
at $T/T_{\rm c0} = 0.58$ and $\mu_{\rm B} H_y/T_{\rm c0} = 1.6$ [the point (i) in Fig.~\ref{fig:PD_LGE}(a)]. The amplitudes of $\Delta_1(x)$ and $\Delta_3(x)$ are almost equivalent and constant, while the phase is almost linearly proportional to the site $i_x$. 
These behaviors in the outer layers are characteristic of the Fulde-Ferrell-type gap function: $\Delta_1(x) \approx \Delta_\textrm{FFLO} e^{i q_0 (x-x_1)}$ and $\Delta_3(x) \approx \Delta_\textrm{FFLO} e^{-i q_0 (x-x_3)}$.
In contrast, the amplitude is modulated with some nodes in $\Delta_2(x)$. The phase shows jumps by $\pm\pi$ at the nodes, 
revealing the Larkin-Ovchinnikov-type gap structure on the inner layer: $\Delta_2(x) \approx \delta \Delta_\textrm{FFLO} 
\cos{q_0 (x-x_2)}$.
These features are consistent with the orbital FFLO superconductivity which occurs mainly on the outer layers. 
This is reasonable because the temperature is below the transition temperature of the orbital FFLO superconductivity and above that of the uniform Ising superconductivity.

\begin{figure}[tb]
  \centering
  \includegraphics[width=85mm]{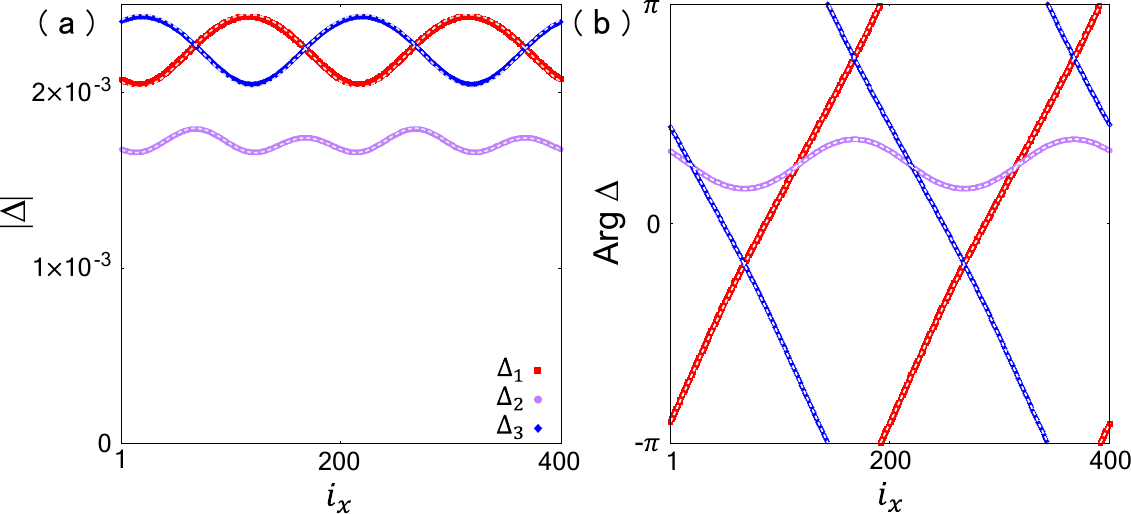}
  \caption{Spatial profile of (a) the amplitude and (b) the phase of the gap functions at $\mu_{\rm B} H_y/ T_{\rm c0} = 1.6$ and $T / T_{\rm c0} = 0.4$. The red, purple, and blue symbols correspond to the top, middle, and bottom layers, respectively. The dashed lines show the fitting by Eq.~\eqref{Gap_form}.
 }
\label{fig:BdG_q0}
\end{figure}

The gap function at a lower temperature $T/T_{\rm c0} = 0.4$ [the point (ii) in Fig.~\ref{fig:PD_LGE}(a)] 
is qualitatively different from Fig.~\ref{fig:BdG_OFFLO}. The amplitude of $\Delta_2(x)$ is nodeless, and the phase oscillates as shown in Fig.~\ref{fig:BdG_q0}. Thus, the gap function on the middle layer represents the "phase-wave" state. 
The gap functions on the outer layers $\Delta_1(x)$ and $\Delta_3(x)$ are also modified from the simple Fulde-Ferrell-type structure, and the amplitudes are largely modulated. 

The superconducting state shown in Fig.~\ref{fig:BdG_q0} can be regarded as a coexistent state of the uniform Ising superconductivity and the orbital FFLO superconductivity. 
As summarized in Table~\ref{Table:LGE_q0_gap}, the Cooper pairs on the top, middle, and bottom layers favor the total momentum $q_0$, 0, and $-q_0$, respectively. Taking into account the corresponding single-$\bm q$ gap functions and the proximity effects of them, the gap functions can be approximated as follows, 
\begin{gather}
\Delta_1(x) = \Delta_\textrm{FFLO} e^{i q_0 (x-x_1)} + \delta \Delta_\textrm{Ising} e^{i \theta} + \delta ^\prime \Delta_\textrm{FFLO} e^{-i q_0 (x-x_3)}, \nn
\Delta_2(x) = \delta \Delta_\textrm{FFLO} \cos{q_0 \left(x - \frac{x_3 + x_1}{2} \right) } e^{i q_0 \frac{x_3 - x_1}{2}} + \Delta_\textrm{Ising} e^{i \theta}, \nn
\Delta_3(x) = \Delta_\textrm{FFLO} e^{-i q_0 (x-x_3)} + \delta \Delta_\textrm{Ising} e^{i \theta} + \delta ^\prime \Delta_\textrm{FFLO} e^{i q_0 (x-x_1)}.
\label{Gap_form}
\end{gather}
As shown in Figs.~\ref{fig:BdG_OFFLO} and \ref{fig:BdG_q0}, the solutions of the BdG equations are fitted well by Eq.~\eqref{Gap_form} with $\Delta_\textrm{FFLO} \gg \delta \Delta_\textrm{FFLO} \gg \delta' \Delta_\textrm{FFLO}$ and $\Delta_\textrm{Ising} \gg \delta \Delta_\textrm{Ising}$. Thus, the superconducting gap functions are represented by a linear combination of the solutions of the linearized gap equation. The components
($\Delta_\textrm{FFLO}, \delta \Delta_\textrm{FFLO}, \delta' \Delta_\textrm{FFLO}$) and
($\Delta_\textrm{Ising}, \delta \Delta_\textrm{Ising}$) are order parameters of the orbital FFLO superconductivity and the uniform Ising superconductivity, respectively. Above $T_{\rm c}^{\rm Ising}$,  $\Delta_\textrm{Ising} = \delta \Delta_\textrm{Ising} =0$ as shown in Fig.~\ref{fig:BdG_OFFLO}. 
In Fig.~\ref{fig:BdG_q0} below $T_{\rm c}^{\rm Ising}$, the coexistent state of orbital FFLO superconductivity and uniform Ising superconductivity is stabilized, which is called layer-selective FFLO state.  The relative phase of the two order parameters $\theta -q_0(x_3-x_1)/2 \simeq \pi/2$ makes $\Delta_2(x)$ essentially complex-valued.
We stress that the layer-selective FFLO phase is absent in bilayer systems since the Ising superconductivity with $q = 0$ does not survive in the high-magnetic-field region.
In general, due to the Fermi surface shift across all layers in the symmetric gauge choice, the layer-selective FFLO state tends to be suppressed in the even-layer systems. 

\begin{figure}[tb]
  \centering
  \includegraphics[width=70mm]{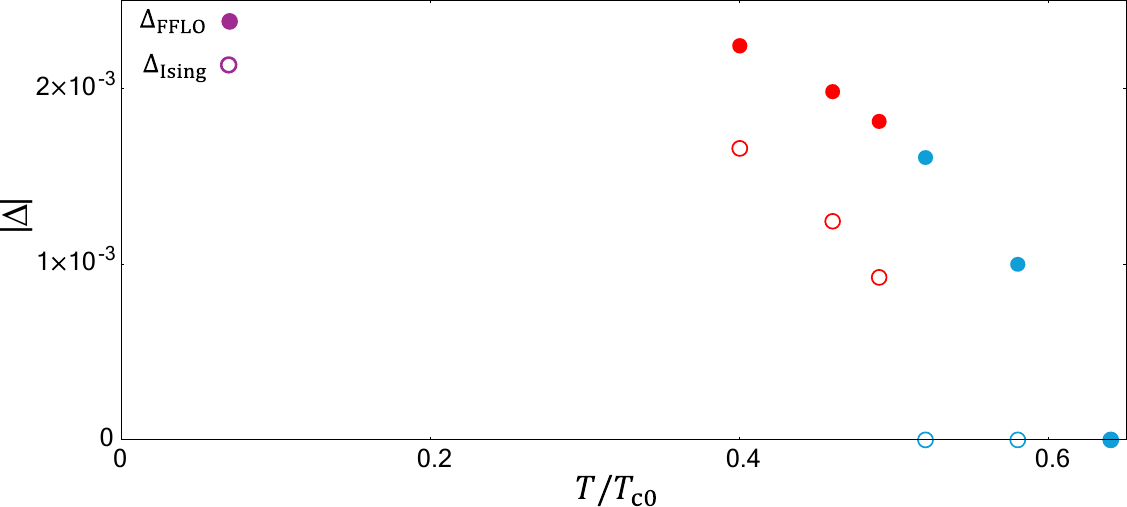}
  \caption{Temperature dependence of the order parameters of orbital FFLO superconductivity $\Delta_{\rm FFLO}$ (closed circles) and uniform Ising superconductivity $\Delta_{\rm Ising}$ (open circles). The magnetic field is set to $\mu_{\rm B} H_y/T_{\rm c0} = 1.6$ as in Figs.~\ref{fig:BdG_OFFLO} and \ref{fig:BdG_q0}. 
  The symbols are highlighted by red (blue) in the layer-selective FFLO (orbital FFLO) phase. 
 }
\label{fig:Tdep}
\end{figure}

\emph{Phase transition.}--- 
The phase transition from the orbital FFLO state to the layer-selective FFLO state is accompanied by symmetry breaking.
When the uniform components are absent ($\Delta_\textrm{Ising}= \delta \Delta_\textrm{Ising} = 0$), the gap functions satisfy
\begin{align}
\Delta_i(x) = e^{i \pi} \Delta_i\left( x + \frac{\pi}{q_0} \right),
\label{Gap_condition}
\end{align}
for all layers. Therefore, the discrete translational symmetry with periodicity $\pi/q_0$ combined with the global phase change 
is preserved in the orbital FFLO state. However, this symmetry is broken in the layer-selective FFLO state because the uniform component is present ($\Delta_\textrm{Ising}, \delta \Delta_\textrm{Ising} \ne 0$). Therefore, these two superconducting phases are distinguished by symmetry. 
We show the temperature dependence of the order parameters in the high-field region $\mu_{\rm B} H_y/T_{\rm c0} = 1.6$ in Fig.~\ref{fig:Tdep}: $\Delta_{\rm FFLO}$ and $\Delta_{\rm Ising}$ are obtained by fitting with Eq.~\eqref{Gap_form}. Below $T_{\rm c}^{\rm Ising}$, the order parameter of uniform Ising superconductivity $\Delta_{\rm Ising}$ smoothly grows, and a discontinuous change in $\Delta_\textrm{FFLO}$ is not observed there. 
Therefore, we conclude the second-order phase transition between the orbital FFLO and layer-selective FFLO phases. 

\begin{figure}[tb]
  \centering
  \includegraphics[width=85mm]{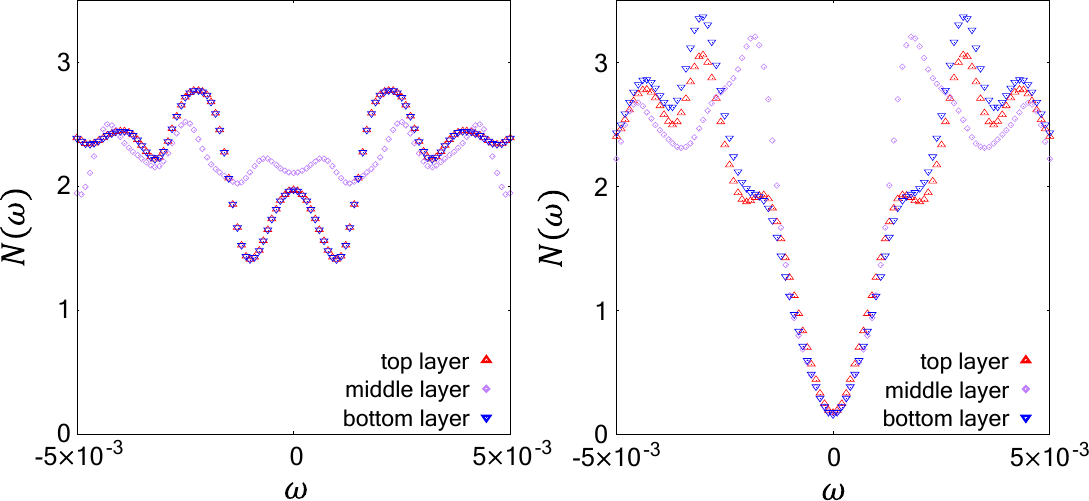}
  \caption{LDOS in (a) the orbital FFLO state at $i_x = 144$, 
  and (b) the layer-selective FFLO state at $i_x = 124$. 
  The parameters in (a) and (b) are the same as Figs.~\ref{fig:BdG_OFFLO} and \ref{fig:BdG_q0}, respectively. The red, purple, and blue symbols represent the top, middle, and bottom layers. 
 }
\label{fig:BdG_LDOS}
\end{figure}

The finite-momentum superconducting phases revealed above are distinguished by the LDOS. 
In the orbital FFLO state, 
we see a large residual LDOS around $\omega = 0$ in all layers [Fig.~\ref{fig:BdG_LDOS}(a)]. However, the LDOS in the layer-selective FFLO state shows a V-shaped structure 
as shown in Fig.~\ref{fig:BdG_LDOS}(b). Since the middle layer in the orbital FFLO state is essentially normal, i.e. only a small gap function is induced by the proximity effect, we can regard this state as a kind of the S-N-S junction. Therefore, the bound states can be formed and give a residual LDOS. Because the LDOS in the outer layers is qualitatively different between the orbital FFLO state and the layer-selective FFLO state, they can be distinguished by spectroscopic measurements such as scanning tunneling microscopy. We find that the LDOS is almost uniform in the orbital FFLO phase, in contrast to the standard Larkin-Ovchinnikov state where the bound states are formed around gap nodes~\cite{Vorontsov2005, Mayaffre2014,Matsuda-Shimahara}. On the other hand, the LDOS in the layer-selective FFLO phase is spatially modulated due to the non-uniform gap magnitudes on the outer layers~\cite{Supplemental}. 

We also find the phase transition between the uniform Ising superconducting phase and the layer-selective FFLO phase by the real-space calculation. The layer-selective FFLO state is stable at $T/T_{\rm c0} = 0.4$ and $\mu_{\rm B} H_y/T_{\rm c0} = 0.8$ [(iii) in Fig.~\ref{fig:PD_LGE}(a)], while the uniform Ising superconducting state is stable in a lower field $\mu_{\rm B} H_y/T_{\rm c0} = 0.6$ [(iv) in Fig.~\ref{fig:PD_LGE}(a)]. The results suggest a first-order phase transition around $\mu_{\rm B} H_y/T_{\rm c0} \simeq 0.7$ consistent with the solution of the linearized gap equation (Fig.~\ref{fig:PD_LGE}) 
and with experimental observation in a thicker NbSe$_2$~\cite{Cao2024}. However, we remark that the finite-size effect can not be avoided in the low-field region because $q_0$ is smaller than that in the high-field region and the finite-size calculation can deal with only discrete values of $q_0$. Therefore, an analysis with larger system size is required to precisely determine the transition line of the uniform Ising superconducting phase and the layer-selective FFLO phase. 

\emph{Discussion.}---
While several experiments suggest the finite-momentum superconducting states in thick samples of NbSe$_2$~\cite{Wan2023, Cho2023, Cao2024, Ji2024},
some experiments in thin samples did not show a signature~\cite{Wan2023, Cao2024}. This seems inconsistent with our results which predict multiple superconducting phases of the orbital FFLO, layer-selective FFLO, and uniform Ising superconducting states in the trilayer NbSe$_2$. However, as is known from the studies of conventional FFLO superconductivity~\cite{Matsuda-Shimahara},  finite-momentum superconductivity is expected to be suppressed by disorders, which may be unavoidable in atomically thin films. 
Thus, we expect realization of the multiple finite-momentum superconducting phases in clean systems, 
which seems consistent with the recent observation of finite-momentum superconductivity in the trilayer NbSe$_2$ blocks in (PbSe)$_{1.14}$(NbSe$_2$)$_3$~\cite{Itahashi2024}.

We emphasize that the mechanism of finite-momentum superconductivity is ubiquitous since it relies only on the layer-dependent Fermi surface shift induced by the orbital effect of an in-plane magnetic field. 
However, the Ising SOC in TMDs is essential because it suppresses the Pauli depairing effect detrimental to the high-field superconducting phases.
Therefore, it is possible to design and control various finite-momentum superconducting phases not only in trilayer NbSe$_2$ but also in various few-layer TMDs. 
The multilayer MoS$_2$ 
is one of the promising candidates because the finite-momentum superconducting state has already been reported in the intercalated bilayer system~\cite{Zhao2023}. 

In summary, we theoretically studied 
the trilayer NbSe$_2$ 
and found two distinct finite-momentum superconducting phases: the (standard) orbital FFLO phase where all the Cooper pairs have finite momentum and the layer-selective FFLO phase where the zero-momentum Cooper pairs and the finite-momentum ones coexist. Our results imply that 
trilayer (or, generally, more than two layer)
TMDs can be platforms for diverse finite-momentum superconductivity.

\begin{acknowledgements}
We are grateful to Y.~M.~Itahashi, Y.~Iwasa, and K.~Kobayashi for fruitful discussions.
This work was supported by JSPS KAKENHI (Grant Nos. JP22H01181, JP22H04933, JP23K22452, JP23K17353, JP23KJ1219, JP24K21530, JP24H00007, JP25H01249).
\end{acknowledgements}

\nocite{*}
\bibliography{reference}

\clearpage


\onecolumngrid

\setcounter{equation}{0}
\setcounter{figure}{0}
\setcounter{table}{0}
\renewcommand{\theequation}{S\arabic{equation}}
\renewcommand{\thefigure}{S\arabic{figure}}
\renewcommand{\thetable}{S\arabic{table}}

\setcounter{secnumdepth}{2}
\renewcommand{\thesection}{\Roman{section}}

\begin{center}
\large{\textbf{
Supplemental Material for \\
"Orbital FFLO and layer-selective FFLO phases in trilayer NbSe$_2$"
}}
\end{center}

\section{DETAILS OF MODEL HAMILTONIAN}
\begin{figure}[b]
  \centering
  \includegraphics[width=90mm]{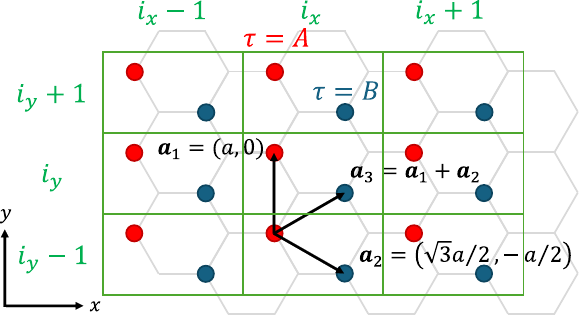}
  \caption{Schematic of the lattice structure in the monolayer NbSe$_2$. The gray line represents the top view of the monolayer NbSe$_2$ and only the sublattice of Nb atoms is explicitly shown. For the Fourier transformation, we introduce additional "sublattice" degree of freedom $\tau = A, B$, and the green lattice corresponds to the "extended" unit cell.
 }
\label{fig:Lattice}
\end{figure}

First, we show the details of the model Hamiltonian for monolayer NbSe$_2$ in the momentum-space representation. Because the Fermi surface mainly consists of Nb $d$ orbitals in the monolayer NbSe$_2$~\cite{Sticlet2019, Wickramantne2020}, it is sufficient to consider only the triangular sublattice of Nb atoms, and we assume a single orbital tight-binding approximation for simplicity. To reproduce the low-energy band structure of the monolayer NbSe$_2$ obtained by \textit{ab initio} calculations \cite{He2018}, $\mathcal{H}_\textrm{kin}$ and $\mathcal{H}_\textrm{Ising}$ in the Hamiltonian include up to the fifth-neighbor and third-neighbor hopping terms, respectively~\cite{Sticlet2019}:
\begin{align}
\xi(\vk) 
=& 2t_1 \left( \cos{2\alpha} + 2 \cos{\alpha} \cos{\beta} \right) 
+ 2t_2 \left( \cos{2\beta} + 2 \cos{3\alpha} \cos{\beta} \right) 
+ 2t_3 \left( \cos{4\alpha} + 2 \cos{2\alpha} \cos{2\beta} \right) \nn
&+ 4t_4 \left( \cos{\alpha} \cos{3\beta} + \cos{4\alpha} \cos{2\beta} + \cos{5\alpha} \cos{\beta} \right) 
+ 2t_5 \left( \cos{6\alpha} + 2 \cos{3\alpha} \cos{3\beta} \right) + \mu ,
\end{align}
\begin{align}
\bm{g}(\vk)
= \left[ 2 \alpha_\textrm{Z1} \left( \sin{2\alpha} + 2 \sin{\alpha} \cos{\beta} \right) \right. 
\left. + 2 \alpha_\textrm{Z2} \left( \sin{4\alpha} + 2 \sin{2\alpha} \cos{2\beta} \right) \right] \hat{z},
\end{align}
where $\alpha = k_y a / 2$ and $\beta = \sqrt{3} k_x a / 2$ with the in-plane lattice constant $a$. All parameters, including interlayer hopping $t_\perp$ and onsite attractive interaction $U$, are summarized in Table~\ref{Table:parameters}. 

\begin{table}[b]
\centering
\begin{tabular}{|c|c|c|c|c|c|c|c|} \hline
 $\mu$ & $t_1$ & $t_2$ & $t_3$ & $t_4$ \\ \hline
 0.023 & 0.0134 & 0.097 & 0.0066 & -0.0102 \\ \hline 
 $t_5$ & $\alpha_\textrm{Z1}$ & $\alpha_\textrm{Z2}$ & $t_\perp$ & $U$ \\ \hline
  -0.0144 & 0.0163 & 0.0013 & 0.012 & 0.156 \\ \hline 
\end{tabular}
\caption{The parameter set (eV) of the model Hamiltonian.}
\label{Table:parameters}
\end{table}

Next, we derive the Fourier-transformed Hamiltonian in the real space, which is used to solve the BdG equation. We begin with the real-space representation with the additional "sublattice" degree of freedom $\tau = A, B$ as shown in Fig.~\ref{fig:Lattice} for later convenience. For example, we can express the nearest-neighbor hopping terms in $\mathcal{H}_\textrm{kin}$ and $\mathcal{H}_\textrm{Ising}$ considering the orbital effect as follows:

\begin{align}
H_\textrm{kin,NN}
= t_1 \sum \{& 
\delta_{\tau, A}( 
\delta_{i_x, i_x^\prime} \delta_{i_y, i_y^\prime-1}  \delta_{\tau^\prime, A} e^{i \vA^{(m)} \cdot \bm{a}_1} 
+ \delta_{i_x, i_x^\prime} \delta_{i_y, i_y^\prime+1} \delta_{\tau^\prime, A} e^{-i \vA^{(m)} \cdot \bm{a}_1} \nn
&~~~~~~+ \delta_{i_x, i_x^\prime} \delta_{i_y, i_y^\prime}  \delta_{\tau^\prime, B} e^{i \vA^{(m)} \cdot \bm{a}_2} 
+ \delta_{i_x, i_x^\prime+1} \delta_{i_y, i_y^\prime-1} \delta_{\tau^\prime,B} e^{-i \vA^{(m)} \cdot \bm{a}_2} \nn
&~~~~~~+ \delta_{i_x, i_x^\prime} \delta_{i_y, i_y^\prime-1}  \delta_{\tau^\prime, B} e^{i \vA^{(m)} \cdot \bm{a}_3}
+ \delta_{i_x, i_x^\prime+1} \delta_{i_y, i_y^\prime} \delta_{\tau^\prime,B} e^{-i \vA^{(m)} \cdot \bm{a}_3} ) \nn
&+ \delta_{\tau, B}( 
\delta_{i_x, i_x^\prime} \delta_{i_y, i_y^\prime-1}  \delta_{\tau^\prime, B} e^{i \vA^{(m)} \cdot \bm{a}_1} 
+ \delta_{i_x, i_x^\prime} \delta_{i_y, i_y^\prime+1} \delta_{\tau^\prime, B} e^{-i \vA^{(m)} \cdot \bm{a}_1} \nn
&~~~~~~+ \delta_{i_x, i_x^\prime-1} \delta_{i_y, i_y^\prime+1}  \delta_{\tau^\prime, A} e^{i \vA^{(m)} \cdot \bm{a}_2} 
+ \delta_{i_x, i_x^\prime} \delta_{i_y, i_y^\prime} \delta_{\tau^\prime,A} e^{-i \vA^{(m)} \cdot \bm{a}_2} \nn
&~~~~~~+ \delta_{i_x, i_x^\prime-1} \delta_{i_y, i_y^\prime}  \delta_{\tau^\prime, A} e^{i \vA^{(m)} \cdot \bm{a}_3} 
+ \delta_{i_x, i_x^\prime} \delta_{i_y, i_y^\prime+1} \delta_{\tau^\prime,A} e^{-i \vA^{(m)} \cdot \bm{a}_3} ) \nn 
&\} c^\dag_{i_x i_y s m \tau} c_{i_x^\prime i_y^\prime s^\prime m^\prime \tau^\prime},
\label{kin_NN}
\end{align}
\begin{align}
H_\textrm{Ising,NN}
= \frac{\alpha_\textrm{Ising1}}{i}  \sum (-1)^{m-1} \{& 
\delta_{\tau, A}( 
\delta_{i_x, i_x^\prime} \delta_{i_y, i_y^\prime-1}  \delta_{\tau^\prime, A} e^{i \vA^{(m)} \cdot \bm{a}_1} 
- \delta_{i_x, i_x^\prime} \delta_{i_y, i_y^\prime+1} \delta_{\tau^\prime, A} e^{-i \vA^{(m)} \cdot \bm{a}_1} \nn
&~~~~~~ + \delta_{i_x, i_x^\prime} \delta_{i_y, i_y^\prime}  \delta_{\tau^\prime, B} e^{i \vA^{(m)} \cdot \bm{a}_2} 
- \delta_{i_x, i_x^\prime+1} \delta_{i_y, i_y^\prime-1} \delta_{\tau^\prime,B} e^{-i \vA^{(m)} \cdot \bm{a}_2} \nn
&~~~~~~- \delta_{i_x, i_x^\prime} \delta_{i_y, i_y^\prime-1}  \delta_{\tau^\prime, B} e^{i \vA^{(m)} \cdot \bm{a}_3}
+ \delta_{i_x, i_x^\prime+1} \delta_{i_y, i_y^\prime} \delta_{\tau^\prime,B} e^{-i \vA^{(m)} \cdot \bm{a}_3} ) \nn
&+ \delta_{\tau, B}( 
\delta_{i_x, i_x^\prime} \delta_{i_y, i_y^\prime-1}  \delta_{\tau^\prime, B} e^{i \vA^{(m)} \cdot \bm{a}_1} 
- \delta_{i_x, i_x^\prime} \delta_{i_y, i_y^\prime+1} \delta_{\tau^\prime, B} e^{-i \vA^{(m)} \cdot \bm{a}_1} \nn
&~~~~~~+ \delta_{i_x, i_x^\prime-1} \delta_{i_y, i_y^\prime+1}  \delta_{\tau^\prime, A} e^{i \vA^{(m)} \cdot \bm{a}_2} 
- \delta_{i_x, i_x^\prime} \delta_{i_y, i_y^\prime} \delta_{\tau^\prime,A} e^{-i \vA^{(m)} \cdot \bm{a}_2} \nn
&~~~~~~- \delta_{i_x, i_x^\prime-1} \delta_{i_y, i_y^\prime}  \delta_{\tau^\prime, A} e^{i \vA^{(m)} \cdot \bm{a}_3} 
+ \delta_{i_x, i_x^\prime} \delta_{i_y, i_y^\prime+1} \delta_{\tau^\prime,A} e^{-i \vA^{(m)} \cdot \bm{a}_3} ) \nn 
&\} c^\dag_{i_x i_y s m \tau} c_{i_x^\prime i_y^\prime s^\prime m^\prime \tau^\prime}.
\label{Ising_NN}
\end{align}
Here we present the summation over the extended unit cell $i^{(\prime)}_x, i^{(\prime)}_y$, spin $s^{(\prime)}$, layer $m^{(\prime)}$, and sublattice $\tau^{(\prime)}$ by the single symbol $\sum$.

Since we consider the magnetic field $\bm{H} = H\hat{y}$ and the corresponding vector potential $\vA = A^{(m)} \hat{x}$, Cooper pairs acquire finite total momentum in the $x$-direction, and we assume that the superconducting state is uniform in the $y$-direction. We then impose periodic boundary conditions along both the $x$- and $y$-directions and obtain the following,

\begin{align}
H_\textrm{kin,NN}
= t_1 \sum \hspace{0cm} ^\prime \{& 
\delta_{\tau, A}( 
2 \cos{k_y a}~ \delta_{i_x, i_x^\prime} \delta_{\tau^\prime, A} \nn
&~~~~~~+ 2 \cos{\frac{1}{2} k_y a}~e^{i \frac{1}{2} k_y a}~e^{i \frac{\sqrt{3}}{2} A^{(m)} a}~\delta_{i_x, i_x^\prime} \delta_{\tau^\prime, B} \nn
&~~~~~~+ 2 \cos{\frac{1}{2} k_y a}~e^{i \frac{1}{2} k_y a}~e^{-i \frac{\sqrt{3}}{2} A^{(m)} a}~\delta_{i_x, i_x^\prime+1} \delta_{\tau^\prime, B} ) \nn
&+ \delta_{\tau, B}( 
2 \cos{k_y a}~ \delta_{i_x, i_x^\prime} \delta_{\tau^\prime, B} \nn
&~~~~~~+ 2 \cos{\frac{1}{2} k_y a}~e^{-i \frac{1}{2} k_y a}~e^{i \frac{\sqrt{3}}{2} A^{(m)} a}~\delta_{i_x, i_x^\prime-1} \delta_{\tau^\prime, A} \nn
&~~~~~~+ 2 \cos{\frac{1}{2} k_y a}~e^{- i \frac{1}{2} k_y a}~e^{- i \frac{\sqrt{3}}{2} A^{(m)} a}~\delta_{i_x, i_x^\prime} \delta_{\tau^\prime, A} ) \nn
&\} c^\dag_{i_x k_y s m \tau} c_{i_x^\prime k_y s^\prime m^\prime  \tau^\prime},
\end{align}
\begin{align}
H_\textrm{Ising,NN}
= \alpha_\textrm{Ising1}  \sum \hspace{0cm} ^\prime (-1)^{m-1} \{& 
\delta_{\tau, A}( 
2 \sin{k_y a}~ \delta_{i_x, i_x^\prime} \delta_{\tau^\prime, A} \nn
&~~~~~~- 2 \sin{\frac{1}{2} k_y a}~e^{i \frac{1}{2} k_y a}~e^{i \frac{\sqrt{3}}{2} A^{(m)} a}~\delta_{i_x, i_x^\prime} \delta_{\tau^\prime, B} \nn
&~~~~~~- 2 \sin{\frac{1}{2} k_y a}~e^{i \frac{1}{2} k_y a}~e^{- i \frac{\sqrt{3}}{2} A^{(m)} a}~\delta_{i_x, i_x^\prime+1} \delta_{\tau^\prime, B} ) \nn
&+ \delta_{\tau, B}( 
2 \cos{k_y a}~ \delta_{i_x, i_x^\prime} \delta_{\tau^\prime, B} \nn
&~~~~~~- 2 \sin{\frac{1}{2} k_y a}~e^{-i \frac{1}{2} k_y a}~e^{i \frac{\sqrt{3}}{2} A^{(m)} a}~\delta_{i_x, i_x^\prime-1} \delta_{\tau^\prime, A} \nn
&~~~~~~- 2 \sin{\frac{1}{2} k_y a}~e^{- i \frac{1}{2} k_y a}~e^{- i \frac{\sqrt{3}}{2} A^{(m)} a}~\delta_{i_x, i_x^\prime} \delta_{\tau^\prime, A} ) \nn 
&\}c^\dag_{i_x k_y s m \tau} c_{i_x^\prime k_y s^\prime m^\prime \tau^\prime},
\end{align}
by the Fourier transformation along the $y$-axis
\begin{align}
c^\dag_{i_x i_y s m \tau} = 
\frac{1}{\sqrt{N_y}} \sum_{k_y} c^\dag_{i_x k_y s m \tau} e^{i k_y i_y}.
\end{align}
Here $\sum \hspace{0cm} ^\prime$ contains the summation over $k_y$ instead of that over $i^{(\prime)}_y$ for $\sum$ in Eqs.~\eqref{kin_NN} and~\eqref{Ising_NN}.

We perform similar Fourier transformation for all the remaining terms in the same way and obtain the gap equation as follows: 
\begin{align}
\Delta_{i_x m \tau} = 
- \frac{U}{N_y} \sum_{k_y} \left\langle c_{i_x -k_y \downarrow m \tau} c_{i_x k_y \uparrow m \tau} \right\rangle.
\end{align}

Note that introducing the additional "sublattice" degree of freedom $\tau$ is not necessary to identify the site, but it is useful for the above Fourier transformation. We can distinguish between the sites ($i_x, A$) and ($i_x, B$) by simply specifying the $x$ position. Therefore, we can define the simple site index as $i^\textrm{site}_x(i_x, \tau) = 2 i_x + \delta_{\tau, B}$ and $i_x$ in the other sections corresponds to $i^\textrm{site}_x$.


\section{CHEBYSHEV POLYNOMIAL METHOD}
We begin with a general $2N \times 2N$ BdG Hamiltonian
\begin{align}
\mathcal{H} = \frac{1}{2} \bm{\Psi}^\dag \hat{H} \bm{\Psi},~~~~
\hat{H} = 
\left(
\begin{array}{cc}
\hat{H}^\textrm{N} & \hat{\Delta} \\
\hat{\Delta}^\dag & (- \hat{H}^\textrm{N})^T 
\end{array}
\right),
\end{align}
where $\hat{H}^\textrm{N}$ is a normal state Hamiltonian matrix and $\hat{\Delta}$ is a superconducting order parameter matrix for the basis
\begin{align}
\bm{\Psi}^\dag = ( \bm{c}^\dag, \bm{c}^T ),~~~~
\bm{c}^{T} = (c_1 , \cdots , c_N).
\end{align}
Here, the index $i$ represents various degrees of freedom, such as site, layer, and spin. The difference between the retarded Green function $\hat{G}^R (\omega)$ and the advanced Green function $\hat{G}^A (\omega)$ is given by
\begin{align}
d_{\alpha \beta}(\omega)~
(= G^R_{\alpha \beta} - G^A_{\alpha \beta} ) 
= - 2 \pi i \sum_\gamma U_{\alpha \gamma} U^*_{\beta \gamma} \delta(\omega - E_\gamma),
\end{align}
where the unitary matrix $\hat{U}$ diagonalizes the BdG Hamiltonian:
\begin{align}
\hat{U}^\dag \hat{H} \hat{U} = \textrm{diag} (E_1, \cdots ,E_{2N}).
\end{align}

To perform the Chebyshev polynomial expansion, we rescale the Hamiltonian and obtain the dimensionless form~\cite{Weisse2006}:
\begin{align}
\tilde{\hat{H}} = (\hat{H} - b)/a,~~~~ \tilde{E}_\gamma = (E_\gamma - b)/a,~~~~ \tilde{\omega} = (\omega - b)/a,
\end{align}
with
\begin{align}
a = (E_\textrm{max} - E_\textrm{min})/2,~~~~
b = (E_\textrm{max} + E_\textrm{min})/2,
\end{align}
where $E_\textrm{max}$ ($E_\textrm{min}$) is larger (smaller) than the maximum (minimum) eigenvalue of the original Hamiltonian 
and all rescaled eigenvalues $\tilde{E}_\gamma$ are restricted to $[ -1, 1]$. If we restrict $\tilde{\omega}$ to $[ -1, 1]$ ($\omega$ to $[E_\textrm{min}, E_\textrm{max}]$), we can apply the Chebyshev polynomial expansion on the delta function in $d_{\alpha \beta}(\omega)$ and obtain~\cite{Nagai2012,Nagai2020}
\begin{align}
d_{\alpha \beta}(\omega) 
= - & \frac{2 \pi i}{a} \sum_\gamma  U_{\alpha \gamma} U^*_{\beta \gamma}
w(\tilde{\omega}) \left(T_0(\tilde{\omega}) T_0(\tilde{E}_\gamma) + 2 \sum_{n = 1}^\infty T_n(\tilde{\omega}) T_n(\tilde{E}_\gamma) ) \right),
\label{domega}
\end{align}
or formally written in the matrix form as
\begin{align}
\hat{d}(\omega) 
= - \frac{2 \pi i}{a} w(\tilde{\omega}) \left(T_0(\tilde{\omega}) T_0(\tilde{\hat{H}}) + 2 \sum_{n = 1}^\infty T_n(\tilde{\omega}) T_n(\tilde{\hat{H}}) ) \right),
\label{domega_mat}
\end{align}
where $T_n (x)$ is the Chebyshev polynomial of the first kind and weight function $w(x) = (\pi \sqrt{1-x^2})^{-1}$ defined with $x \in [-1,1]$, which satisfy the following relations~\cite{Weisse2006}:
\begin{align}
T_0(x) = 1,~~ T_1(x) = x,~~ T_{n+1}(x) = 2 x T_n(x) - T_{n-1}(x),
\label{rec_CP}
\end{align}
and 
\begin{align}
\int^1_{-1} w(x) T_n(x) T_m(x) dx
= \frac{1 + \delta_{n,0}}{2} \delta_{n,m},
\end{align}
or it can be written in an explicit form:
\begin{align}
T_n (x) = \cos[n \arccos(x)].
\end{align}

We can calculate the superconducting mean field by using Eq.~\eqref{domega_mat}. Using $\hat{d}(\omega)$, the mean field is written as
\begin{align}
\left\langle c_i c_j \right\rangle
= - \frac{1}{2 \pi i} \int^\infty_{-\infty} d\omega f(\omega)  \bm{e}(j)^T \hat{d}(\omega) \bm{h}(i),
\label{SCMF}
\end{align}
where $f(\omega)$ is the Fermi distribution function and $\bm{e}(i)$ and $\bm{h}(i)$ are the 2$N$-dimensional vector defined as 
\begin{align}
\left[ \bm{e}(i) \right]_\gamma = \delta_{i, \gamma},~~~~
\left[ \bm{h}(i) \right]_\gamma = \delta_{i+N, \gamma}.
\end{align}
Inserting Eq.~\eqref{domega_mat}, we obtain
\begin{align}
\left\langle c_i c_j \right\rangle 
= \sum_{n = 0}^\infty \mathcal{T}_n \bm{e}(j)^T \bm{h}_n(i),
\label{SCMF_CPM_sup}
\end{align}
where
\begin{gather}
\mathcal{T}_n = \frac{2}{1+\delta_{n,0}} \int^1_{-1} d\tilde{\omega} f(a \tilde{\omega} + b) w(\tilde{\omega}) T_n(\tilde{\omega}),
\label{Tn_CPM_sup} \\
\bm{h}_n(i) = T_n(\tilde{\hat{H}}) \bm{h}(i).
\end{gather}
Reflecting the recurrence relation of the Chebyshev polynomial [Eq.~\eqref{rec_CP}], $\bm{h}_n(i)$ satisfies
\begin{gather}
\bm{h}_0(i) = \bm{h}(i),~~
\bm{h}_1(i) = \tilde{\hat{H}} \bm{h}(i),~~
\bm{h}_{n+1}(i) = 2 \tilde{\hat{H}} \bm{h}_n(i) - \bm{h}_{n-1}(i).
\end{gather}
We note that Eq.~\eqref{Tn_CPM_sup} can be calculated by using the Chebyshev–Gauss quadrature: 
\begin{align}
\mathcal{T}_n \approx 
\frac{\pi}{N_\textrm{CG}} \sum_{i=1}^{N_\textrm{CG}} f(a \tilde{\omega}_i + b) T_n(\tilde{\omega}_i),
\end{align}
where
\begin{align}
\tilde{\omega}_i = \cos(\frac{(2i-1) \pi}{2 N_\textrm{CG}}).
\end{align}

We can also calculate the local density of states~\cite{Nagai2020}
\begin{align}
N_\alpha (\omega) = - \frac{1}{\pi} \Im G^R_{\alpha \alpha} (\omega) = - \frac{1}{2 \pi i} d_{\alpha \alpha} (\omega).
\end{align}
Because $d_{ii}(\omega) = \bm{e}(i)^T \hat{d}(\omega) \bm{e}(i)$, we obtain
\begin{align}
N_i (\omega) = \sum_{n = 0}^\infty \mathcal{T}^\prime_n (\tilde{\omega}) \bm{e}(i)^T \bm{e}_n(i),
\label{LDOS_CPM_sup}
\end{align}
by using Eq.~\eqref{domega_mat}
and
\begin{gather}
\mathcal{T}^\prime_n (\tilde{\omega}) = \frac{2}{a(1+\delta_{n,0})} w(\tilde{\omega}) T_n(\tilde{\omega}), \nn
\bm{e}_n(i) = T_n(\tilde{\hat{H}}) \bm{e}(i).
\end{gather}
In addition to $\bm{h}_n(i)$, $\bm{e}_n(i)$ also satisfies the following relation:
\begin{gather}
\bm{e}_0(i) = \bm{e}(i),~~
\bm{e}_1(i) = \tilde{\hat{H}} \bm{e}(i),~~
\bm{e}_{n+1}(i) = 2 \tilde{\hat{H}} \bm{e}_n(i) - \bm{e}_{n-1}(i).
\end{gather}
We emphasize that these expressions can be calculated without any diagonalization.  

Although Eqs.~\eqref{SCMF_CPM_sup} and~\eqref{LDOS_CPM_sup} are exact only when we take the infinite sum of the Chebyshev polynomial, these equations are approximately correct with relatively small cutoff $N_{\rm c}$,
\begin{gather}
\left\langle c_i c_j \right\rangle 
\approx \sum_{n = 0}^{N_{\rm c}} g_n \mathcal{T}_n \bm{e}(j)^T \bm{h}_n(i), \nn
N_i (\omega) \approx \sum_{n = 0}^{N_{\rm c}} g_n \mathcal{T}^\prime_n (\tilde{\omega}) \bm{e}(i)^T \bm{e}_n(i).
\end{gather}
Here we introduce the Jackson kernel 
\begin{align}
g_n = \frac{1}{N_{\rm c} + 2} 
\left( (N_{\rm c} + 2 - n) \cos{\frac{\pi n}{N_{\rm c} + 2}} + \sin{\frac{\pi n}{N_{\rm c} + 2}} \cot{\frac{\pi}{N_{\rm c} + 2}} \right),
\end{align}
for better convergence with a small cutoff~\cite{Weisse2006}.

\begin{figure}[h]
  \centering
  \includegraphics[width=55mm]{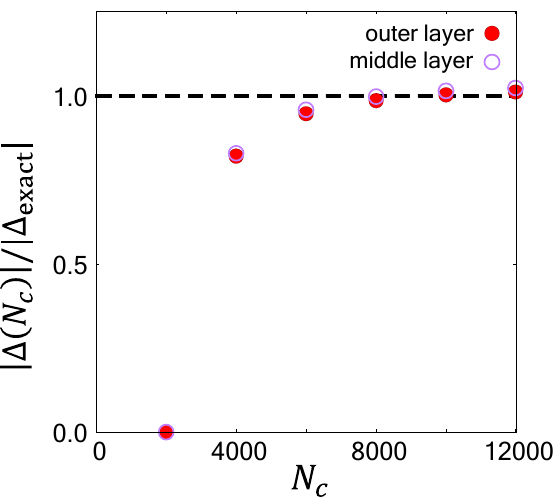}
  \caption{Cutoff dependence of the gap amplitude obtained by the Chebyshev polynomial method at $\mu_{\rm B} H_y/ T_{\rm c0} = 1.6$ and $T / T_{\rm c0} = 0.4$. We assume the spatially uniform gap function and compare its magnitude to $\Delta_\textrm{exact}$ calculated by the standard approach in the momentum space with the same parameter set. We see the convergence for the cutoff $N_{\rm c}$.
 }
\label{fig:Ncdep}
\end{figure}

We check the accuracy of the calculation by reproducing the spatially uniform gap obtained by the calculation in the momentum space. The cutoff dependence of the superconducting gap obtained by solving the BdG equation in the real space 
is shown in Fig.~\ref{fig:Ncdep}. We choose $N_{\rm c} = 10000$ for all calculations in this study.

\section{GAP FUNCTIONS IN THE LOW-FIELD REGION AND THE FIRST-ORDER PHASE TRANSITION}
\begin{figure}[tb]
  \centering
  \includegraphics[width=100mm]{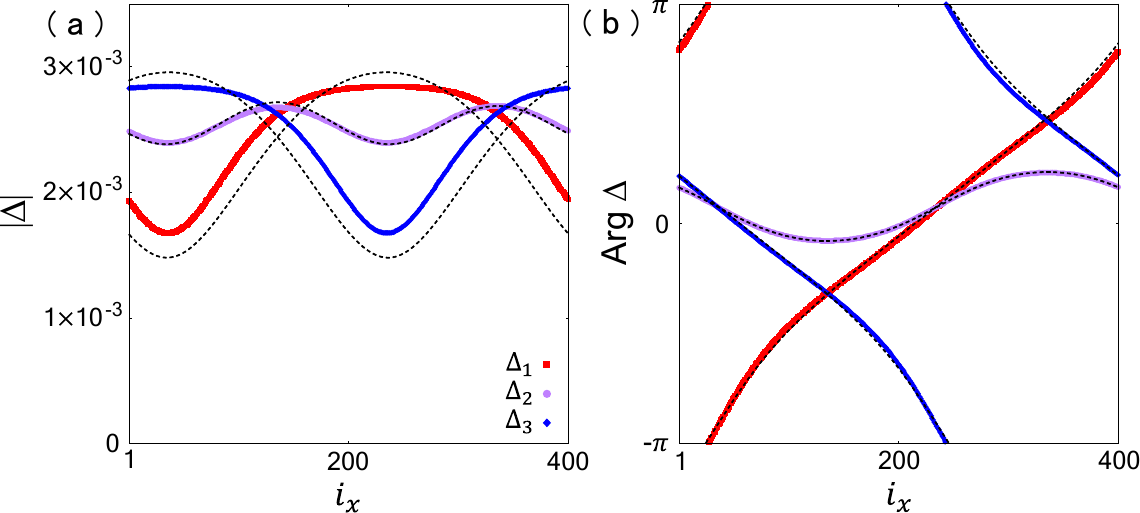}
  \caption{Spatial profile of (a) the amplitude and (b) the phase of the gap functions at $\mu_{\rm B} H/ T_{\rm c0} = 0.8$ and $T / T_{\rm c0} = 0.4$. The red, purple, and blue symbols correspond to the top, middle, and bottom layers, respectively. The dashed lines are the fitting by Eq.~\eqref{Gap_form}.
 }
\label{fig:H0.8}
\end{figure}

We show the amplitude and the phase of the gap functions at $T/T_\textrm{c0} = 0.4$ and $\mu_\textrm{B} H /T_\textrm{c0} = 0.8$ in Fig.~\ref{fig:H0.8} [the point (iii) in Fig.~\ref{fig:PD_LGE}(a)]. They show the periodic structure with a period twice as long as that at $\mu_\textrm{B} H /T_\textrm{c0} = 1.6$, as expected from Fig.~\ref{fig:PD_LGE}(b). The difference between the solution of the BdG equation and the fitting by Eq.~\eqref{Gap_form} is larger than that at $\mu_\textrm{B} H /T_\textrm{c0} = 1.6$ shown in Figs.~\ref{fig:BdG_OFFLO} and \ref{fig:BdG_q0}. This result suggests that the higher-harmonic components with $q = 2 q_0, 3 q_0, \cdots$ are not negligible in the low-field and low-temperature region. We confirmed that the nonuniform state with the finite-$q$ gap functions in Fig.~\ref{fig:H0.8} is more stable than the uniform state by calculating the free energy.  

\begin{figure}[tb]
  \centering
  \includegraphics[width=100mm]{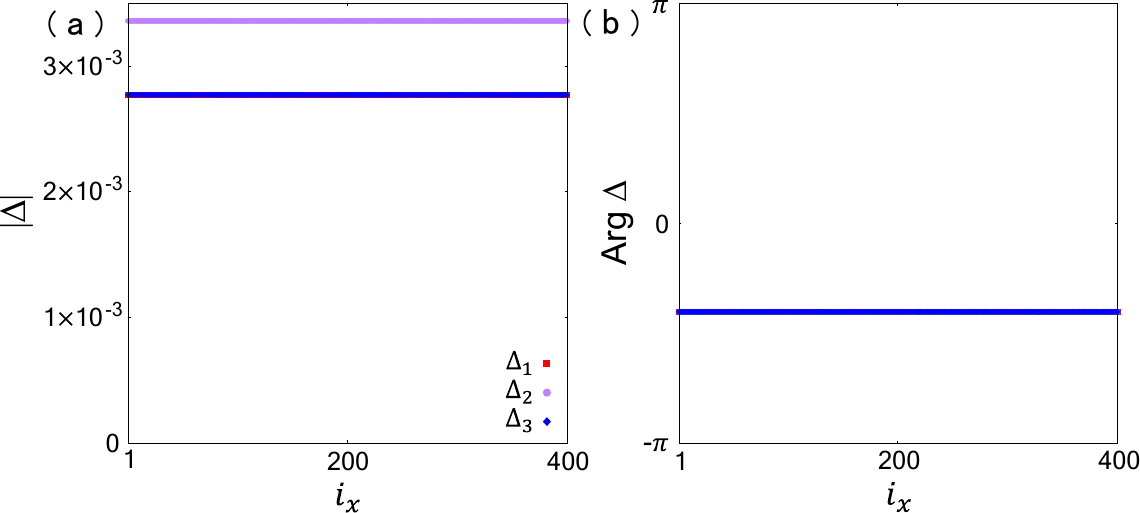}
  \caption{Spatial profile of (a) the amplitude and (b) the phase of the gap functions at $\mu_{\rm B} H/ T_{\rm c0} = 0.6$ and $T / T_{\rm c0} = 0.4$. The red, purple, and blue symbols correspond to the top, middle, and bottom layers, respectively. The amplitudes
on the two outer layers are almost equivalent. The phases are nearly the same across all layers.
 }
\label{fig:H0.6}
\end{figure}

On the other hand, the uniform gap function is obtained at a lower magnetic field $\mu_\textrm{B} H /T_\textrm{c0} = 0.6$ as shown in Fig.~\ref{fig:H0.6} [the point (iv) in Fig.~\ref{fig:PD_LGE}(a)]. Comparing Fig.~\ref{fig:H0.6}(a) with Fig.~\ref{fig:H0.8}(a), we see that the finite-$q$ component suddenly disappears and the uniform component on the outer layers seems to rapidly develop at the phase transition. We thus conclude that a first-order transition occurs between the uniform Ising superconducting phase and the layer-selective FFLO phase around $\mu_\textrm{B} H /T_\textrm{c0} \approx 0.7$.



\section{SITE DEPENDENCE OF LDOS}
\begin{figure}[tb]
  \centering
  \includegraphics[width=100mm]{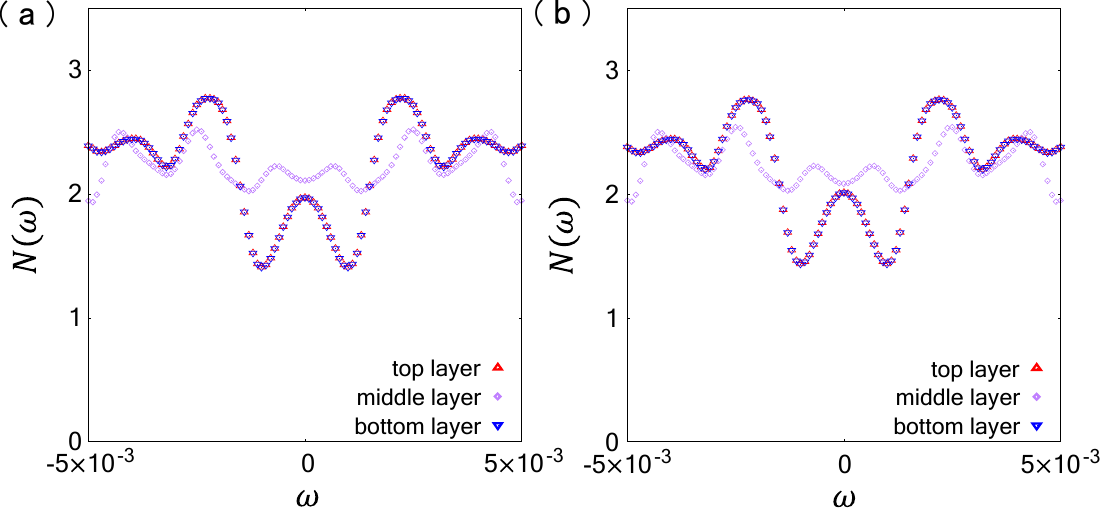}
  \caption{Local density of states (LDOS) in the orbital FFLO phase at $\mu_{\rm B} H_y/ T_{\rm c0} = 1.6$ and $T / T_{\rm c0} = 0.58$. We show the LDOS on the two sites (a) $i_x = 144$ and (b) $i_x = 94$. The red, purple, and blue symbols correspond to the top, middle, and bottom layers, respectively. 
 }
\label{fig:LDOS_sitedep_OFFLO}
\end{figure}

We show the LDOS in the orbital FFLO state at $\mu_{\rm B} H_y/ T_{\rm c0} = 1.6$ and $T / T_{\rm c0} = 0.58$ in Fig.~\ref{fig:LDOS_sitedep_OFFLO}. We calculate the LDOS at $i_x = 144$ [Fig.~\ref{fig:LDOS_sitedep_OFFLO}(a)] where $\Delta_2(x)$ has the maximum value and at $i_x = 94$ [Fig.~\ref{fig:LDOS_sitedep_OFFLO}(b)] which is the node of $\Delta_2(x)$ [see Fig.~\ref{fig:BdG_OFFLO}(a)]. The LDOS is almost independent of the site, although we see a slightly larger DOS around $\omega = 0$ at the gap node.

\begin{figure}[tb]
  \centering
  \includegraphics[width=140mm]{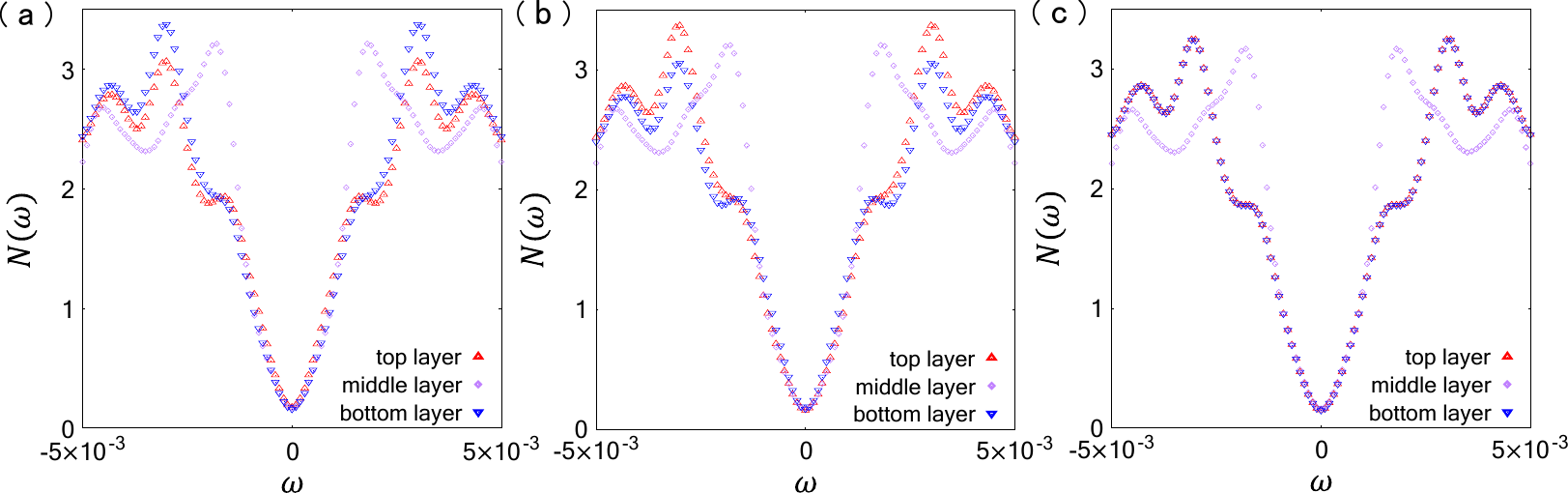}
  \caption{LDOS in the layer-selective FFLO phase at $\mu_{\rm B} H_y/ T_{\rm c0} = 1.6$ and $T / T_{\rm c0} = 0.4$. The sites are (a) $i_x = 124$, (b) $i_x = 212$, and (c) $i_x = 168$. The red, purple, and blue symbols correspond to the top, middle, and bottom layers, respectively. 
 }
\label{fig:LDOS_sitedep_q0}
\end{figure}

On the other hand, we can see the site dependence of the LDOS in the layer-selective FFLO state. Figure~\ref{fig:LDOS_sitedep_q0}(a) shows the higher peak around $\omega = 3 \times 10^{-3}$ on the bottom layer at $i_x = 124$, while it is on the top layer at $i_x = 212$ as shown in Fig.~\ref{fig:LDOS_sitedep_q0}(b). This behavior is related to the site dependence of the gap amplitude. At $i_x = 124$, $\Delta_1(x)$ has the maximum value and $\Delta_3(x)$ has the minimum value, while the relation is reversed at $i_x = 212$ [see Fig.~\ref{fig:BdG_q0}(a)]. At $i_x = 168$, where the amplitudes of $\Delta_1(x)$ and $\Delta_3(x)$ are almost the same, the LDOS is also nearly equivalent.

\end{document}